# Scaling laws for doublet craters formed by low-velocity impacts of unequal-mass spheres into a granular bed

Haruto Kitagawa,[1] Ririha Obara,[1] and Yu Matsuda[1,*]

[1]Department of Modern Mechanical Engineering, Waseda University, 3-4-1 Ookubo, Shinjuku-ku, Tokyo, Japan.

**ABSTRACT**. Understanding the formation mechanism of doublet craters is an important challenge for advancing knowledge in astronomy and granular physics. In this study, we investigated the craters formed by low-velocity impacts of two steel spheres with different masses into a granular bed. Even when the masses of the spheres were different, a figure-eight-shaped doublet crater was formed similarly to the case with spheres with equal masses, and a central ridge was observed near the center. However, it was found that the resulting shape became asymmetric even in the absence of a time delay between the impacts. Experimental results showed that the total length of the doublet crater increased depending on the spacing between the two spheres and the ratio of their impact energies. These experimental results followed a theoretical model derived from the scaling law in which the crater diameter is proportional to the 1/4 power of the impactor's kinetic energy. A theoretical model was also developed to describe the overlap of the two craters, and it was shown that the overlap increases as the two spheres become closer. It was also shown that the diameters of the individual craters vary depending on the time difference between the impacts. Especially, the crater diameter formed by the second impacting sphere tends to be larger than that predicted by the scaling law due to the effect of fluidization induced by the first sphere impact.

## I. INTRODUCTION.

Craters are characteristic landforms observed across a wide range of scales [1,2]. The most familiar example of craters is those formed by raindrops, [3-5] while the most representative examples are craters found on the surfaces of planetary bodies such as Mars, Earth, and the Moon. [6-8] Because the formation mechanism of craters involves multiple complex phenomena, it is difficult to establish an appropriate physical model. Therefore, many studies have analyzed craters using small-scale and simple experimental setups in which spherical objects are dropped onto granular beds. Based on these studies, it has been both theoretically and empirically derived that, in the case of solid spheres, the crater diameter is proportional to the one-fourth power of the kinetic energy just before impact [9-11]. It is also known that when the impactor is a liquid drop or an elastic body, the exponent in the above scaling law becomes smaller than 1/4 [1,9,12,13]. In addition, various studies have been conducted on factors such as temperature variation in the granular bed [14], slope inclination [15,16], and wet granular bed [17], and knowledge regarding craters formed by single impactors has gradually been consolidated.

There are craters known as doublet craters, which are formed when two meteoroids strike the surface at nearly the same time and only a short distance apart. It is reported that the observed number of plausible doublet craters is 2–4% on Earth and 2–3% on Mars. [18] Studies on doublet craters have been conducted, including experimental investigations of the penetration of two rod-shaped objects into granular beds at small scales, [19] as well as analyses of morphological changes using numerical simulations [18]. More recently, Jiménez-Valdez et al. [20] conducted drop experiments in which two steel spheres of equal mass were released onto a granular bed, and demonstrated that the resulting crater shape varies depending on the spacing between the spheres and the time difference in their release. The total length of a doublet crater varies significantly depending on the distance between the two spheres. When the spheres are very close to each other, the resulting crater is equivalent to that formed by a single sphere with twice the impact energy. When the impacts occur simultaneously, the central ridge—one of the characteristic features of a doublet crater—has a straight shape, and it becomes increasingly curved as the time delay between the impacts increases. It was also observed that the sphere impacting later penetrates deeper into the granular bed than the one impacting first. Since the sphere that impacts later penetrates more deeply, the shape of the crater it forms may differ from that formed by the earlier-impacting sphere. Furthermore, doublet craters formed by two spheres with different masses—and therefore different impact energies—remain an area open to further investigation.

In this study, we performed experiments involving the low-velocity impact of two steel spheres with different masses onto a granular bed. The resulting crater dimensions were obtained using a three-dimensional (3D) profilometer, and the impact process was captured via high-speed imaging. Based on the measurement data, we investigated the effects of time delay, inter-sphere distance, and energy ratio on the total length and overlap of the resulting doublet craters, and derived relevant scaling laws.

*Contact author: y.matsuda@waseda.jp

## II. MODEL

The scaling laws for the total length $L_1$ of a doublet crater and the length of the overlapping region $L_2$, as illustrated in Fig. 1, are considered. Since $L_1$ represents the overall size of the doublet crater and $L_2$ indicates the extent of overlap between the two individual craters, these parameters serve as effective indicators for evaluating the crater morphology.

As described in Introduction, the relationship between the crater diameter $D_c$ and the impactor's kinetic energy $E$ is given by

$$D_c = kE^\alpha. \qquad (1)$$

where $k$ is a constant, and the exponent $\alpha$ is equal to 1/4 for a rigid sphere impact. Here, we consider a time delay between the impacts of the two spheres. Then, the relationships between the crater diameter $D_f$ and the kinetic energy $E_f$ of the first (earlier) impact, and between the crater diameter $D_s$ and the kinetic energy $E_s$ of the second (later) impact, are expressed as

$$D_f = kE_f^\alpha, \qquad (2)$$
$$D_s = kE_s^\alpha. \qquad (3)$$

The energy ration between these spheres $\varepsilon$ is defined as

$$\varepsilon = \frac{E_s}{E_f}. \qquad (4)$$

Here, based on geometric considerations, the total length $L_1$ can be expressed in terms of $D_f$, $D_s$, and the center-to-center distance between the two spheres $\delta$ as follows:

$$L_1 = \frac{1}{2}(D_f + D_s) + \delta. \qquad (5)$$

Substituting Eqs. 2, 3, and 4 to Eq. 5, the total length $L_1$ is written as

$$L_1 = \frac{1}{2}kE_f^\alpha(1 + \varepsilon^\alpha) + \delta. \qquad (6)$$

Eq. 6 can be nondimensionalized in two different ways: when nondimensionalized by $D_f$, it takes the following form as

$$\frac{L_1}{D_f} = \frac{1}{2}(1 + \varepsilon^\alpha) + \frac{\delta}{D_f}; \qquad (7)$$

when nondimensionalized by $D_f + D_s$, it takes the form as

$$\frac{L_1}{D_f + D_s} = \frac{1}{2} + \frac{\delta}{D_f + D_s}. \qquad (8)$$

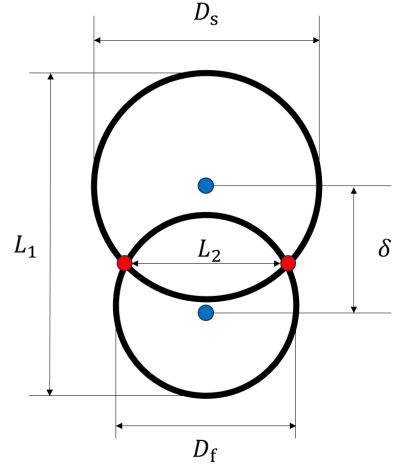

FIG 1. Definition of characteristic lengths in a doublet crater

Next, the length of the overlapping region $L_2$ is expressed as follows,

$$L_2 = \sqrt{D_f^2 - \left(\frac{D_f^2 - D_s^2}{4\delta}\right)^2}. \qquad (9)$$

Substituting Eqs. 2, 3, and 4 to Eq. 9, Eq. 9 is represented as

$$L_2 = kE_f^\alpha\sqrt{1 - (kE_f^\alpha)^2\left(\frac{1 - \varepsilon^{2\alpha}}{4\delta}\right)^2}. \qquad (10)$$

This equation is also nondimensionalized as

$$\frac{L_2}{D_f} = \sqrt{1 - (kE_f^\alpha)^2\left(\frac{1 - \varepsilon^{2\alpha}}{4\delta}\right)^2}. \qquad (11)$$

## III. METHODS

In this study, glass beads ($2.5 \text{ g/cm}^3$; diameter: $214 \pm 37 \text{ μm}$; ASONE, Japan) were used as the granular medium. The beads were poured into a petri dish (diameter: 122 mm; depth: 26 mm) and the surface was leveled. The packing fraction was set to approximately 61% for all experiments. Steel spheres made of high-carbon chromium bearing steel (SUJ2) with diameters of $2.49 \pm 0.02$, $2.99 \pm 0.02$, $3.99 \pm 0.02$, $4.49 \pm 0.02$, $4.99 \pm 0.02$, and $5.99 \pm 0.02$ mm were used as impactors. An electromagnetic holder (TMEH-B3, TRUSCO NAKAYAMA, Japan) was used to control the release of the spheres. Each sphere was placed in a hole on an A5052 aluminum plate attached to the bottom of the electromagnetic holder. When the power supply to the holder was turned off, the spheres dropped freely onto the granular bed. The release height, defined as the distance from the surface of the granular bed to the bottom of the electromagnetic holder, was fixed at $H = 163.5 \pm 0.05$ mm. The center-to-center distance $\delta$ between the two spheres was controlled by adjusting the hole spacing on the aluminum plate. The time difference $\Delta t$ between the impacts was adjusted by utilizing the residual magnetization retained in the steel spheres.

*Contact author: y.matsuda@waseda.jp

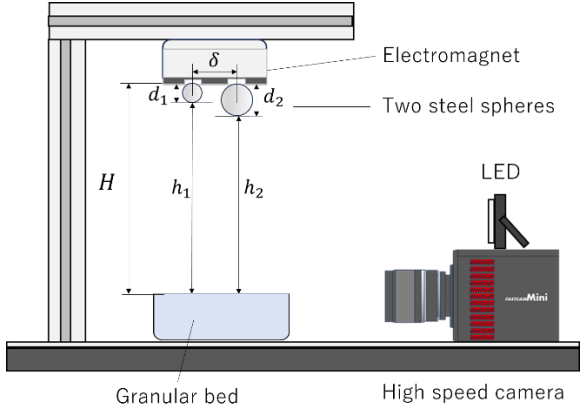

FIG 2. Schematic illustration of the experimental apparatus.

TABLE I. Combinations of impactor diameters and the corresponding values of center-to-center distance $\delta$ and time delay $\Delta t$ in the experiments.

| Pairs of sphere diameters $(d_1, d_2)$ [mm] | center-to-center distance $\delta$ [mm] | Time delay of the impacts $\Delta t$ [ms] |
|---|---|---|
| (2.5, 2.5) | 5 – 7.5 | -3.889 – 3.056 |
| (2.5, 3) | 5 – 8.8 | -11.67 – 3.889 |
| (3, 4) | 6 – 10 | -3.611 – 4.444 |
| (4.5, 6) | 10 – 15 | -5.278 – 6.944 |
| (5, 5) | 10 – 15 | -4.444 – 3.889 |
| (5, 6) | 10 – 15 | -7.500 – 6.111 |

The moment of impact was recorded using a high-speed camera (FASTCAM Mini AX50, Photron, Japan) at a frame rate of 3600 fps with an exposure time of 1/3600 s. The surface profile of the resulting craters was measured using a 3D laser displacement sensor (LJ-S040, KEYENCE, Japan). The crater diameters were calculated by fitting circles to the peak edges of the crater rim extracted from the obtained 3D data. A schematic of the experimental setup is shown in Fig. 2. The time delay of the impacts $\Delta t$ was defined as the time interval between the larger sphere contacting the bed surface and the smaller sphere making contact. The combinations of spheres used, along with the corresponding values of $\delta$ and measured $\Delta t$, are summarized in Table 1.

## IV. RESULTS AND DISCUSSION

We first examine the crater diameter formed when a single sphere is dropped onto the granular bed to validate our experimental method. Fig. 3 shows the relationship between the common logarithm of the impact energy, $\log_{10} E$, and the common logarithm of the resulting single crater diameter, $\log_{10} D_c$, for six types of steel spheres with different diameters. As shown in Fig. 3, the experimental results follow the known scaling law $D_c \propto E^{1/4}$.

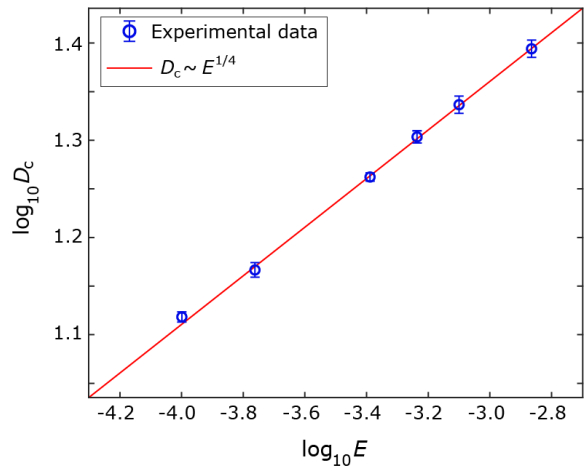

FIG 3. Relations between single crater diameter $D_c$ and impactor kinetic energy $E$.

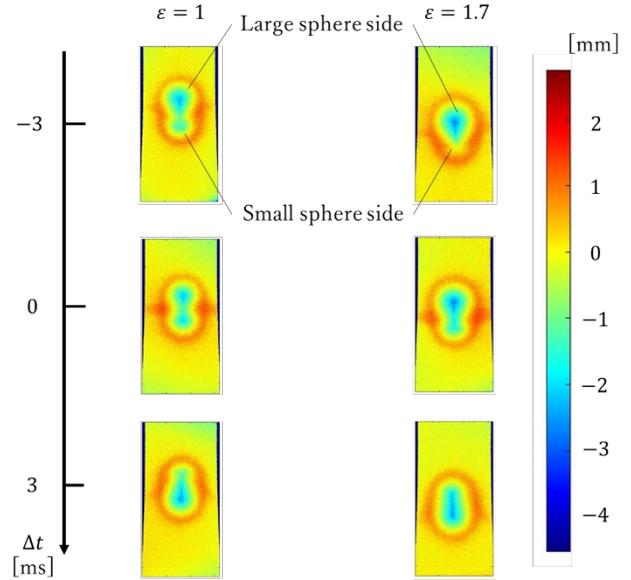

FIG 4. Morphological variations of doublet craters as different $\Delta t$ and $\varepsilon$. The original height of the granular surface is denoted as 0 mm.

*Contact author: y.matsuda@waseda.jp

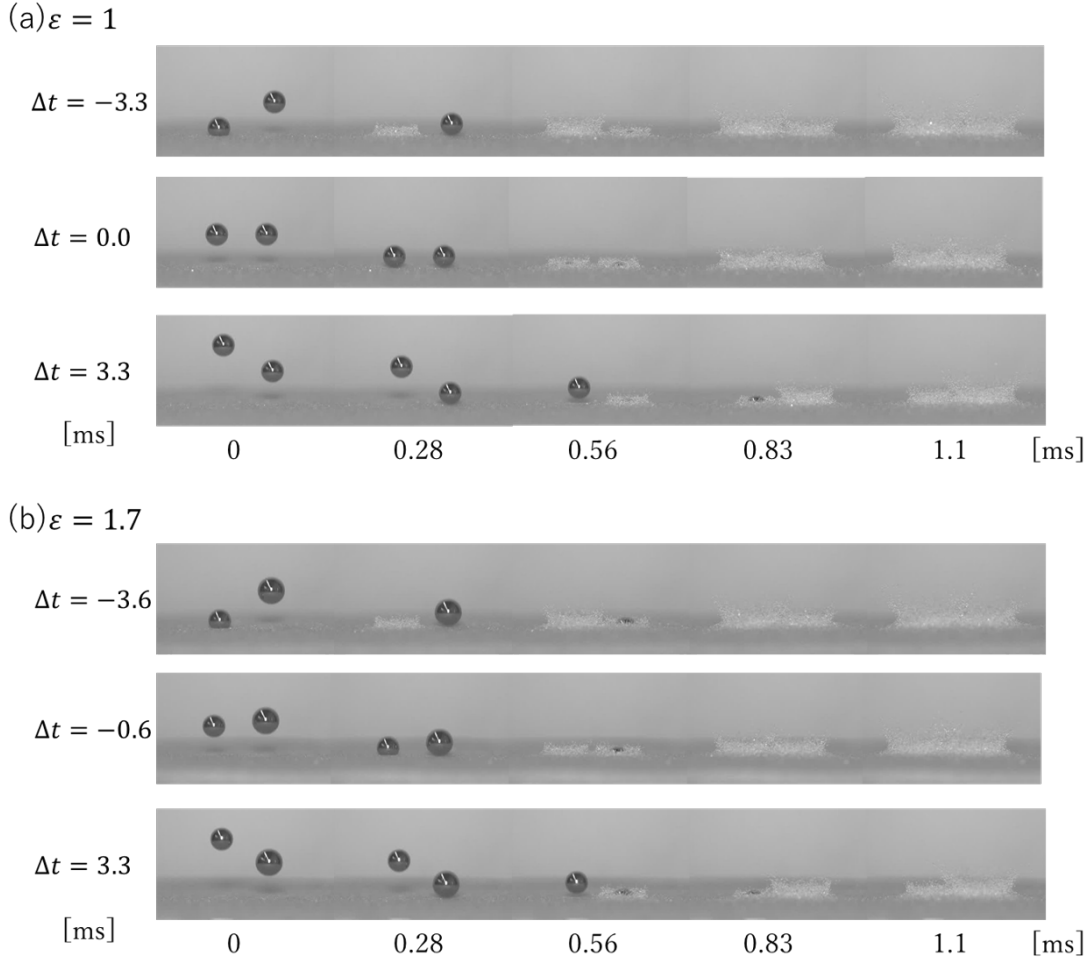

FIG 5. Impact behavior of two spheres onto a granular bed for (a) $\varepsilon = 1$ and (b) $\varepsilon = 1.7$.

Then, the formation of doublet crater was investigated. The height profiles of doublet craters acquired using a 3D laser displacement sensor are presented in Fig. 4. The high-speed images of the impact are shown in Fig. 5. These figures show representative cases for $\varepsilon = 1$ and 1.7 with $\Delta t = -3$, 0, and 3 ms. In Jiménez-Valdez et al. [20], it was shown that in impacts involving equal-mass spheres, the shape of the doublet crater strongly depends on the center-to-center distance $\delta$, and the central ridge becomes increasingly curved as the time delay $\Delta t$ increases. As seen in Fig. 4, a similar dependence on $\Delta t$ is observed even in doublet craters formed by spheres of unequal mass, indicating that the curvature of the central ridge still varies with the time delay. When $\Delta t \cong 0$ ms, the crater is symmetric in the equal-mass case, but an asymmetric shape is observed when there is an energy difference between the spheres.

In the discussion of the total length $L_1$, $L_1$ was nondimensionalized using $D_f^s$ and $D_s^s$ instead of $D_f$ and $D_s$, because $D_f$ and $D_s$ influence each other: the crater formed by the first impact is partially eroded by the later impact, while the crater formed by the later impact is affected by the presence of the first. Using $D_f^s$ and $D_s^s$, which are obtained from single-sphere impacts under the same conditions, simplifies the discussion. Fig. 6 shows the relation between the center-to-center distance $\delta$ and the total length $L_1$, both nondimensionalized by $D_f^s$. As shown in Fig. 6, the experimental data is well represented by Eq. 7. These results indicate that the use of $D_f^s$ for normalization is appropriate. Fig. 7 presents the experimental results of $\delta$ and $L_1$ normalized by $D_f^s + D_s^s$. These results also agree with the theoretical model described by Eq. 8, showing that the data collapse onto a single linear relationship regardless of scale and energy ratio. However, in the experiments conducted by Jiménez Valdez et al. [20], it was reported that Eq. 7 does not hold when the two spheres are placed very close to each other—within a region where they can be approximated as a single point mass with the combined mass of both spheres. In their study, the nondimensionalization of $\delta$ was performed using the particle diameter $d_p$, and it was shown that when $\delta/d_p = 1$, the resulting crater diameter matches that formed by a single sphere with twice the mass. In the present study, the experiments were conducted in a region of $\delta/d_p > 1$. As a result, such convergence to a single-mass behavior was not observed.

*Contact author: y.matsuda@waseda.jp

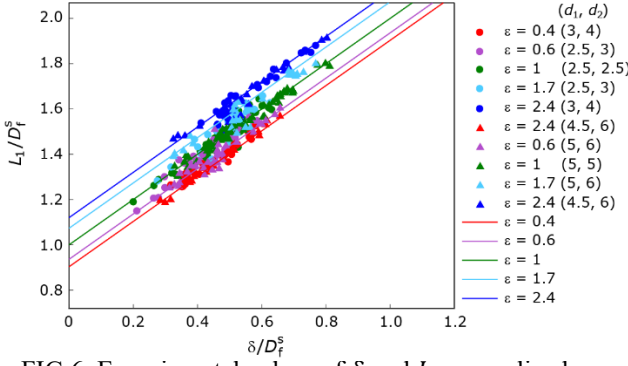

FIG 6. Experimental values of $\delta$ and $L_1$ normalized by $D_f^s$

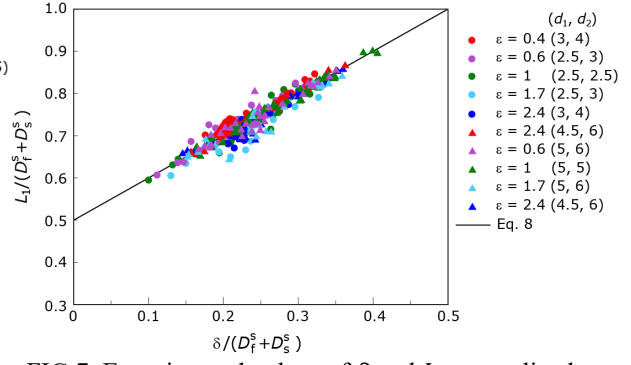

FIG 7. Experimental values of $\delta$ and $L_1$ normalized by $D_f^s + D_s^s$

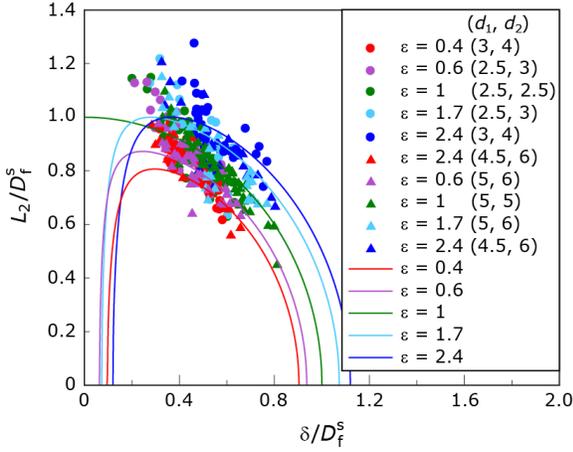

FIG 8. Relation between $\delta$ and $L_2$ normalized by $D_f^s$. The curves of Eq. 11 are also shown.

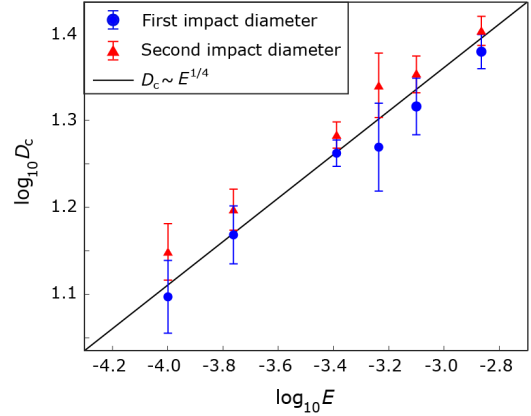

FIG 9. Relations between single crater diameter $D_c$ and impactor kinetic energy $E$ for first and second impact.

We also compared the theoretical model with the experimental results for $L_2$. Fig. 8 shows the relation between $\delta$ and $L_2$, both normalized by $D_f^s$. As shown in Fig. 9, $L_2$ decreases as $\delta$ increases. Moreover, $L_2$ increases with increasing energy ratio $\varepsilon$. These results indicate that the overlap between the two craters becomes larger when the spheres are closer and when the energy ratio is higher, which is consistent with both theoretical predictions and experimental observations. For $\delta/D_f^s < 0.2$, two spheres are very close, and the resulting craters are almost fully merged; data were not obtained in this region. For $\delta/D_f^s > 1$, the two craters are separated from each other, and a doublet crater is not formed. While the overall trend in Fig. 8 generally follows the theoretical model given by Eq. 11, some data points at $0.2 < \delta/D_f^s < 0.4$ exhibit deviations from the model. To investigate these deviations, we focused on the change in crater diameter due to the time delay between impacts. The relations between the crater diameters of first and second impact and the impact energy are shown in Fig. 9. As shown in Fig. 9, it can be observed that the crater diameter formed by the second impacting sphere tends to be larger than that predicted by the scaling law in Eq. 1. To explore this further, we considered the influence of bed fluidization caused by the impact of the first sphere. Fig. 10 shows the relationship between the nondimensionalized separation distance $\delta/(v_m|\Delta t|)$ and the crater diameter ratio $D_s^d/D_f^d$, where $D_f^d$ and $D_s^d$ denote the diameters of the individual craters formed in the doublet crater configuration. The velocity $v_m$ represents the propagation speed of fluidization within the granular bed, and it is considered to be on the same order as the impact velocity of the sphere in this study.

*Contact author: y.matsuda@waseda.jp

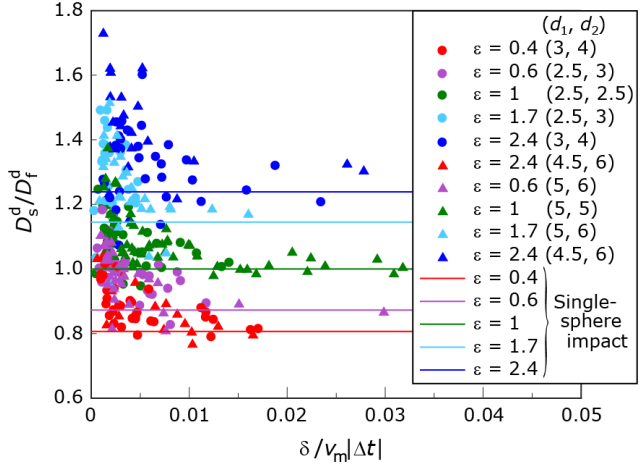

FIG 10. Relationship between nondimensionalized sphere separation $\delta/(v_\mathrm{m}|\Delta t|)$ and crater diameter ratio $D_s^d/D_f^d$

Then, the nondimensional parameter $\delta/(v_\mathrm{m}|\Delta t|)$ indicates whether the impact point of the second sphere, at the moment of its collision, lies within the region influenced by the fluidization caused by the first impact. In Fig. 10, the five horizontal lines represent the single-impact crater diameter ratios for each energy ratio. The plot reveals that as $\delta/(v_\mathrm{m}|\Delta t|)$ increases, the crater diameter ratio $D_s^d/D_f^d$ decreases and approaches the corresponding value for single-sphere impacts. This indicates that the second sphere forms a larger crater when it impacts within the region influenced by the fluidization induced by the first impact. Furthermore, it can be observed that for all energy ratios, the crater diameter ratio converges to that of single-sphere impacts when $0.01 < \delta/(v_\mathrm{m}|\Delta t|) < 0.015$. In other words, this range can be interpreted as the critical separation distance beyond which the influence of fluidization induced by the first impact no longer affects the second impact.

## V. CONCLUSIONS

In this study, we investigated how the energy ratio between two falling steel spheres affects the morphology of doublet craters formed in a granular bed. It was shown that even in the absence of a time delay, an energy difference between the two spheres results in an asymmetric crater shape. The crater morphology was found to depend strongly on the distance between the two spheres. The total length of the doublet crater increases linearly with the separation distance and also increases with increasing energy ratio, as demonstrated both theoretically and experimentally. Moreover, the overlap between the two individual craters comprising the doublet crater becomes larger as the separation distance decreases. When the separation distance is fixed, a higher energy ratio leads to a greater overlap, in agreement with theoretical predictions and experimental results. Discrepancies between the theoretical model and experimental values were attributed to variations in crater diameter arising from time delays between impacts. These variations were found to be particularly significant when the separation distance was small enough for the granular bed to be fluidized by the impact of the first sphere.

*Contact author: y.matsuda@waseda.jp

*Contact author: y.matsuda@waseda.jp